\documentclass[aps,reprint,amsmath,amssymb,amsfonts,showpacs,nofootinbib]{revtex4-1}
\usepackage{graphicx}
\usepackage{hyperref}
\usepackage{mathrsfs}
\usepackage{color}
\usepackage{placeins}
\usepackage[T1]{fontenc}
\usepackage{bib_macro}
\usepackage[version=3]{mhchem}
\newcommand*{\fontpackage}{}
\renewcommand*{\fontpackage}{%
 txfonts%
}
\newcommand*{\fontpackageoptions}{}
\renewcommand*{\fontpackageoptions}{
}
\newcommand*{\isomathoptions}{}
\renewcommand*{\isomathoptions}{%
OMLmathsans,%
 sfdefault=zavm
}
\usepackage[\fontpackageoptions]{\fontpackage}
\usepackage[\isomathoptions]{isomath}
\usepackage{esint}  
\usepackage{bbm}
\newcommand{\sou}{\text{S}}
\newcommand{\obs}{\text{O}}

\newcommand{\Omm}{\Omega_\text{m}}
\newcommand{\rhom}{\bar\rho_\text{m}}
\newcommand{\rhol}{\rho_\Lambda}
\newcommand{\upPhi}{\mathrm{\Phi}}
\newcommand{\upPsi}{\mathrm{\Psi}}
\newcommand{\lapl}{\upDelta}
\newcommand{\NM}{\mathrm{NM}}
\newcommand{\Long}{\mathrm{L}}
\newcommand{\Newt}{\mathrm{N}}
\newcommand{\Lambdaup}{\mathrm{\Lambda}}

\newcommand{\bn}{{\vec n}}

\newcommand{\de}{\delta}
\newcommand{\De}{\upDelta}
\newcommand{\Om}{\Omega}

\newcommand{\be}{\begin{equation}}
\newcommand{\ee}{\end{equation}}
\newcommand{\bea}{\begin{eqnarray}}
\newcommand{\eea}{\end{eqnarray}}
\newcommand{\bean}{\begin{eqnarray*}}
\newcommand{\eean}{\end{eqnarray*}}
\newcommand{\dd}{\partial}

\newcommand{\HH}{{\mathcal H}}

\renewcommand{\rho}{\varrho}
\usepackage{upgreek}

\renewcommand{\rho}{\varrho}

\renewcommand*{\vec}{\vectorsym}

\begin{document}
\title{Newtonian N-body simulations are compatible with cosmological perturbation theory}
\author{$^{1,2}$Thomas Haugg}
\email{thomas.haugg@physik.lmu.de}
\author{$^{1,2}$Stefan Hofmann}
\email{stefan.hofmann@physik.lmu.de}
\author{$^{1,2,3}$Michael Kopp}
\email{michael.kopp@physik.lmu.de}
\affiliation{$^{1}$Arnold Sommerfeld Center for Theoretical Physics,Ludwig-Maximilians-Universit\"at,   \\
Theresienstr. 37, 80333 Munich, Germany   \\
$^{2}$Excellence Cluster Universe, Boltzmannstr. 2, 85748 Garching, Germany \\
$^{3}$University Observatory, Ludwig-Maximillians University Munich,  \\ Scheinerstr. 1, 81679 Munich, Germany } 

\begin{abstract}
Contrary to recent claims \cite{FS12}, Newtonian N-body simulations of collisionless Dark Matter in a $\Lambdaup$CDM background
are compatible with general relativity and
are not invalidated by general relativistic effects at the linear level. 	
This verdict is based on four facts. (1) The system of linearized Einstein equations and conservation laws 
is well-posed in the gauge invariant formulation and physically meaningful.
(2) Comparing general relativity with its Newtonian approximation at a given order 
in perturbation theory is only meaningful at the level of observables.
(3) The dynamics of observables describing a dust fluid in general relativity 
and its Newtonian approximation agree at the linear level. 
Any disagreement for observables on the lightcone are well-known, of which the most dominant is gravitational lensing.
(4) Large fluctuations in the Hubble parameter contribute significantly only to gravitational lensing effects.
Therefore, these fluctuations are not in conflict with 
Newtonian N-body simulations beyond what has already been carefully taken into account
using ray tracing technology.
\end{abstract}

\keywords{Backreaction, Newtonian N-body simulation}
\pacs{98.80.Cq}
\maketitle

\section{Introduction}\label{sec:intro}

There seems to be little doubt that the gravitational formation of structures on scales deep inside the Hubble volume
is accurately described by Newtonian gravity, with general-relativistic effects entering at subleading level
in the perturbative description. 
Nevertheless it was claimed \cite{FS12} recently that effects in cosmological perturbation theory
become dominant over $2$nd order effects in the Newtonian approximation on scales larger than $10\,$Mpc.  

In greater detail the argument was based on the following. The evolution of cosmological perturbations was calculated
in a particular coordinate system, called the {\it Newtonian matter gauge} (NM), in which the linear density contrast
$\delta_\NM$ and the peculiar velocity $v_\NM$ coincide with the corresponding quantities $\delta_\Newt=\delta_\NM$, $v_\Newt=v_\NM$
in the Newtonian (N) approximation. It was found that fluctuations in the local Hubble parameter 
$\delta^{H}_{\NM}\equiv-1-K_\NM/(3H)$, where $K_\NM$ denotes the extrinsic curvature,
are of order $\mathcal{O}(\delta_\NM)$.   
The significance of this observation was evaluated by comparing $\delta^{H}_{\NM}$ to the size of
second order corrections $\delta_\Newt^{(2)}$ in the Newtonian approximation. 
For comoving scales $k$ and redshifts $z$ with
\begin{align}
\delta^{H}_{\NM}\geq \delta_\Newt^{(2)}\qquad\mathrm{and}\qquad \delta_\Newt^{(2)}\ll \delta_\Newt <1\; ,\label{comp}
\end{align}
it was argued that relativistic effects linear in cosmological fluctuations dominate over nonlinear Newtonian effects 
well within the domain of validity of perturbation theory.
Based on this criterion, the authors of \cite{FS12} found that linear cosmological perturbations 
on an Einstein-de Sitter background dominate over Newtonian nonlinearities for scales
$k^{-1}>10\,\mathrm{Mpc}$ during the redshift interval $z\in[0.4,750]$, from which they concluded
that Newtonian N-body simulations cannot be trusted on these scales during the specified redshift interval.

Since by choice of gauge, $\delta_\Newt=\delta_\NM$, $v_\Newt=v_\NM$, while the fluctuations in the Hubble parameter
are absent in the Newtonian approximation, and thus in N-body simulations, it is interesting to ask how
these fluctuations become manifest in observables? It might be expected \cite{FS12} that fluctuations in the Hubble parameter
cause additional redshift space distortions, because the observed redshift depends on 
$K_\NM$ integrated along the line of sight between observer and source, 
in addition to the peculiar velocities of observer and source, $v_{\NM}^{\mathrm{O}}$ and $v_{\NM}^{\mathrm{S}}$,
respectively. 
We find indeed that these fluctuations contribute considerably to redshift space distortions, however,
only via gravitational lensing, which is well known as {\it lensing magnification}, e.g~\cite{G67,KS87,YFZ09,Y10,M00,BD11,CL11}.
These lensing induced distortions are taken into account in N-body simulations using ray tracing technology
\cite{WCO98,HHWS09}. 

Cosmological perturbation theory can be compared to its Newtonian approximation in a meaningful way only 
by comparing observables. Observables are by definition gauge invariant combinations of the 
perturbations. The set of observables should be specified before a choice of gauge is implemented,
after the gauge redundancies have been removed it is impossible to identify observables.
However, once the observables have been identified, gauge freedom is not sacrosanct and can either be removed
or, equivalently, used to rewrite the theory using gauge invariant variables. Both procedures are perfectly valid.
A comparison of observables using gauge invariant perturbations has been performed in \cite{NH12} and was 
criticized in \cite{FS12} as follows:
``The initial conditions must be specified on a spatial Cauchy hypersurface, which in the context of cosmological perturbation theory corresponds to a particular foliation of space-time, i.e., to a hypothetical observer who is able to determine physical quantities on a spatial hypersurface. The relativistic-Newtonian correspondence mixes the quantities defined on different spatial hypersurfaces and thus no hypothetical observer in the Einsteinian world could actually determine these combined quantities.''
\begin{figure}[t]
\centering
\includegraphics[width=0.45 \textwidth]{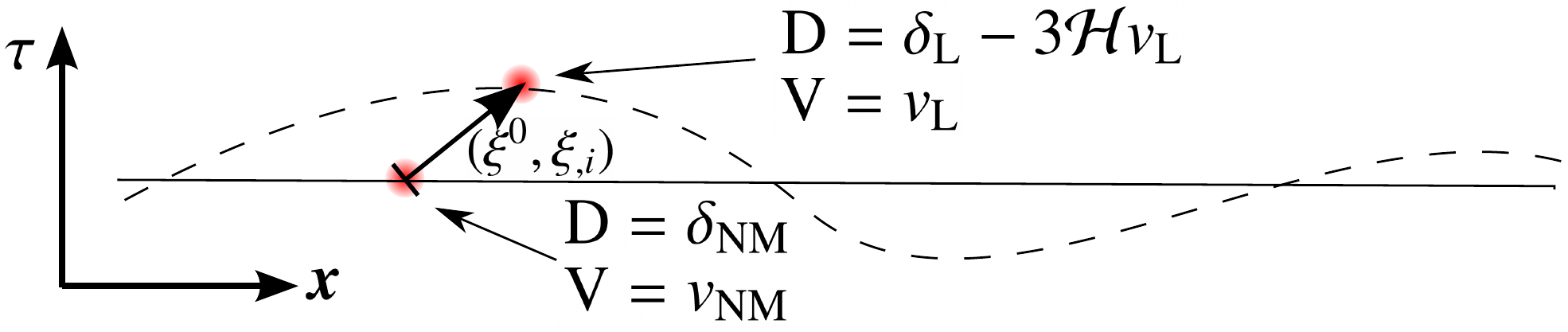}
\caption{D and V agree on different hypersurfaces ({\it full} Newton matter gauge and {\it dashed} longitudinal gauge) evaluated at the same coordinate values in their respective coordinate systems. Only the physical interpretation of D and V changes and on both hypersurfaces an hypothetical observer could determine D and V.}
\label{fig:hyper}
\end{figure}

This is a misconception that requires immediate clarification. 
Although gauge invariant variables might have a simple physical interpretation only in one specific gauge,
they present observables in all other coordinate systems and, in particular, different hypersurfaces, as well.
For instance, let D be a gauge invariant variable which reduces to the linear density contrast
$\delta_\sou$ in synchronous (S) gauge, measured by an observer at rest with respect to synchronous coordinates.
Let $V$ denote a gauge invariant variable which reduces to the peculiar velocity $v_\Long$ as measured
by an observer at rest relative to the longitudinal coordinate system. This is the situation referred
to ``defined on different spatial hypersurfaces'' in the above quote. However, (D,V) are defined
in all possible coordinate systems and observers adopted to different and arbitrary coordinate systems
can measure D and V and will find the same numerical values for them. By construction, it is not the definition 
of gauge invariant variables that is tied to certain hypersurfaces, but it is their physical interpretation. 
This is why any smart observer will construct the set of observables before choosing a particular gauge 
to measure them, because even if not all observables have a convenient physical interpretation in 
the observer's coordinate frame, they still resemble the only meaningful quantities for any other observer.
The only reasonable academic debate between observers adopted to different coordinate systems is about
the physical interpretation of the gauge invariant variables. 

As an example, consider an observer who is adopted to a longitudinal (L) coordinates system. This observer
will interpret D as D$_\Long=\delta_\Long-3 \HH v_\Long$. Although the observer has a physical 
interpretation for the gauge dependent quantities $(\delta_\Long,v_\Long)$, (s)he understands the 
necessity to construct gauge invariant combinations involving $(\delta_\Long,v_\Long)$, rather than
assuming any other observer adopted to an arbitrary coordinate system would agree on the values of the gauge dependent
quantities. The observer adopted to the longitudinal frame can measure $(\delta_\Long,v_\Long)$
on a hypersurface, set up the initial value problem for an appropriate 
evolution equation involving (D,V) and finally solve for them, instead of 
$(\delta_\Long,v_\Long)$, see Fig.\,\ref{fig:hyper}.

The plan for the rest of this paper is as follows. In Section \ref{sec:Hypobs} it is shown explicitly 
that the hypothetical observables $(\delta_\Newt,v_\Newt)$ of linearized Newtonian gravity 
and (D,V) of linearized Einsteinian gravity obey the same evolution equations
on a flat $\Lambdaup$CDM background (in \cite{FS12} only $\Lambdaup=0$ has been considered).
This was expected since (D,V) reduce to $(\delta_\NM,v_\NM)$ for observers at rest
in the Newtonian matter coordinate system.

However, a real observer cannot measure (D,V) on a hypersurface, because (s)he can only observe
the light cone associated with (her) him via light rays, which get affected by curvature perturbations.   
This induces a feedback of general relativistic effects on (D,V), although (D,V) still coincide
with $(\delta_\NM,v_\NM)$ on each hypersurface. Note that the status of (D,V) as observables
is not challenged by the practical obstacles that prevent any observer from measuring 
them on the entire hypersurface. 
 
Section \ref{sec:LCobs} is devoted to investigate the impact of Hubble parameter fluctuations 
on the linear density contrast observed along a light cone. This is a well-known example
highlighting the fact that $\delta^{H}_{\NM}$ contributes significantly only to gravitational lensing. 
The strongest additional redshift space distortions due to gravitational lensing  
\cite{G67,KS87,YFZ09,Y10,M00,BD11,CL11} originate indeed from $\delta^{H}_{\NM}$
(as well as from intrinsic curvature, see Section \ref{sec:LCobs}).
The gauge invariant lensing term, of course, does not depend on the gauge the observer preferred. 
An observer adopted to the longitudinal gauge finds a negligible contribution from 
$\delta^{H}_{\Long}$ to gravitational lensing. 

Our main result is that Newtonian N-body simulations are in congruence with 
cosmological perturbation theory and are not threatened by relativistic effects
at the linear level although relativistic effects can become significant. We give a pratical dictionary to use N-body simulation data to evaluate these corrections. 

\section{Congruence of linear observables on a hypersurface}\label{sec:Hypobs}
In this section we show that a pressureless fluid in a universe with $\Lambdaup$CDM background
geometry can be characterized by observables (D,V) in $1$st-order cosmological perturbation theory 
that obeys evolution equations identical to those governing the evolution of $(\delta_\Newt,v_\Newt)$.

Restricting attention solely to scalar perturbations, the conformal evolution equations 
for $(\phi_\Newt,\delta_\Newt,v_\Newt)$
in the Newtonian approximation are given by 
\begin{subequations}\label{Napt}
\bea
\lapl \phi_\Newt&=& \tfrac{3}{2} \HH^2 \Omm \de_\Newt \; ,\\
\delta_\Newt'+\lapl v_\Newt &=&0 \; ,\\
v_\Newt'+\HH v_\Newt&=&- \phi_\Newt \; ,
\eea
\end{subequations}
with  $\HH$ denoting the conformal expansion rate of the background determined
by Friedmann equations
\begin{subequations}\label{Freeq}
\bea
3 \HH^2&=& 8\pi G (\rhom+\rhol)a^4 \;, \\
4\pi G a^2 \rhom&=&\tfrac{3}{2}\Omm \HH^2=a^2(\HH^2-\HH ') \; ,
\eea
\end{subequations}
where $\rhom\propto a^{-3}$ and $\Omm$ denotes the background matter density
relative to the critical density. 

The Newtonian perturbation variables ($\delta_\Newt,v_\Newt$) are defined 
by the dark matter density $\rho_\mathrm{m}= \rhom (1+\delta_\Newt)$
and the peculiar velocity $\vec{v}_\Newt= \nabla v_\Newt$. 
The triple $(\phi_\Newt,\delta_\Newt,v_\Newt)$ constitutes the set of observables
relevant for the discussion.

The corresponding description in general relativity requires a background metric around which 
the geometry fluctuates. Restricting attention again solely to scalar metric fluctuations,
$(\phi,w,\psi,h)$,
the total metric field is give by 
\bea \label{metric}
{\rm d}s^2 
&=&
a^2\Big[-(1+2\phi)\,{\rm d}\tau\otimes{\rm d}\tau +2w_{,i}\, {\rm d}\tau \otimes {\rm d}x^i +  \\
&&\quad ~+[(1-2\psi)\de_{ij}+ 2h_{,ij} ]{\rm d}x^i\otimes{\rm d}x^j\Big]~.\nonumber 
\eea
The total pressureless 
source is encoded in $T=\rho_\mathrm{m} U\otimes U$, with $\rho_\mathrm{m}= \rhom (1+\delta)$
and $U=(1-\phi, \nabla v)/a$.
Altogether the dynamical degrees of freedom $(\phi,\psi,w,h,\delta,v)$
are the gauge dependent metric, density and velocity perturbations. 

Observables can be constructed by the following gauge invariant linear combinations:
\begin{subequations}\label{Ginvv}
\bea
\upPhi &=& \phi + {[(w-h^\prime)a]}^\prime /a \;, \\
\upPsi&=& \psi - \HH(w-h') \; ,\\
\text{D}&=&\de-3 \HH(v+w) \; ,\\
\text{V}&=& v+h'\; .
\eea
\end{subequations}
The perturbed Einstein and conservation equations then yield evolution
equations \cite{NH12} for the gauge invariant quantities 
$(\upPhi,\upPsi,\text{D},\text{V})$:
\begin{subequations}\label{cpt}
\bea
\lapl \upPhi&=& \tfrac{3}{2} \HH^2 \Omm\text{D}\; ,\\
\text{D}'+\lapl \text{V} &=&0 \; ,\\
\text{V}'+\HH \text{V}&=&- \upPhi \; ,
\eea
\end{subequations}
where the background equations \eqref{Freeq} and the linearized $(0j)$- and $(ij)$-Einstein equations
have been used: 
\begin{subequations}\label{momcon}
\bea
\HH \upPhi+ \upPsi'&=& -\tfrac{3}{2} \HH^2 \Omm\text{V} \;  ,\\
\upPhi&=&\upPsi  \;.
\eea
\end{subequations}
Comparing (\ref{cpt}) with the Newtonian approximation (\ref{Napt}) it is evident
that the evolution equations are identical in form. 
In additions, $(\upPhi,\text{D},\text{V})$ constitute the triple of observables relevant
for this discussion. Of course, any linear combination of these gauge invariant variables
constitutes an equally legitimate observable. The triple $(\upPhi,\text{D},\text{V})$
is favored only to establish directly the correspondence between 
relativistic observables and observables in the Newtonian approximation at the linear level. 

Note that $(\upPhi,\upPsi)$ has the same quasi-static dynamics as $\phi_\Newt$,
which allow us to
qualify relativistic corrections to Newtonian observables, e.g.~\eqref{Dez} below, as large or small
in comparison to $2$nd order corrections in the Newtonian approximation \eqref{comp}.

Let us emphasize again that there is no gauge $\mathcal{G}$ in which simultaneously $$\upPhi=\phi_\mathcal{G},~~\upPsi=\psi_\mathcal{G},~~\text{D}=\de_\mathcal{G},~~\text{V}=v_\mathcal{G}\; .$$
Different observers simply assign different physical meaning to these gauge invariant variables. 
For instance, in longitudinal, synchronous and Newtonian matter gauge the following 
physical interpretations hold:\\

\begin{tabular}{|c|c|c|c|c|c|}
\hline \hline Gauge & \;&$\upPhi$ & $\upPsi$ & D &V \\  \hline
L &\;&  $\phi_\Long$ & $\psi_\Long$& $\delta_\Long-3 \HH v_\Long$ &$v_\Long$\\ 
S &\;&  $-h_\sou'' -h_\sou'\HH$ & $\psi_\sou+h_\sou' \HH$& $\delta_\sou$ &$h_\sou'$\\ 
NM &\; & $-{v'}_{\NM}-v_\NM\HH$&$\psi_\NM+v_\NM \HH$ &$\de_\NM$& $v_\NM$\\ \hline \hline 
\end{tabular}
\section{Linear observables on the lightcone} \label{sec:LCobs}
A physical observer is in practice not able to measure (D,V) everywhere on any hypersurface.
Instead, a physical observer can only learn about (D,V) by employing light rays traveling along 
(her) his respective light cone. As a consequence of such an observation campaign, 
(D,V) become subject to relativistic effects that have no Newtonian counterpart.
The light rays will be gravitationally lensed and these lensing effects will be attributed to
(D,V). Since gravitational lenses are absent in the Newtonian approximation, the dictionary  
$(\upPhi,\text{D},\text{V})\leftrightarrow(\phi_\Newt,\delta_\Newt,v_\Newt)$ is challenged. 

As an example, consider the linear density fluctuation $\De(\bn,z)$ at the observed redshift $z$
and direction $-\vec{n}$ on the sky or, equivalently, in the direction $\vec{n}$ of the incoming
light ray at the observer's space-time position $(\tau_\obs,\vec{x}_\obs)$, 
within the relativistic framework of $1$st order cosmological perturbation theory.
It is given by the gauge invariant expression \cite{BD11}
\bea \label{Dez}
\De(\bn,z) &=& \nonumber \text{D}-\frac{1}{\HH}\dd_r^2  \text{V} - \frac{1}{r_S}\int_0^{r_S}\hspace{-3mm}d\lambda \frac{r_S-r}{r}\lapl_\Om(\upPhi+\upPsi) + \\ &&~+ 
 \left(\frac{\HH'}{\HH^2}+\frac{2}{r_S\HH}\right)\left(\upPsi-\dd_r \text{V}+ 
 \int_0^{r_S}\hspace{-3mm}d\lambda(\upPhi+\upPsi)'\right) 
  \\  \nonumber  &&~   
+\frac{1}{\HH}\upPhi'  +3 \HH \text{V} -2 \upPhi + \upPsi+\frac{2}{r_S}\int_0^{r_S}\hspace{-3mm}d\lambda  (\upPhi+\upPsi)~,  \nonumber
\eea
where all functions are evaluated along the unperturbed light cone $\vec{x}=\vec{x}_\obs- \vec{n}r(z)$,  
$\tau= \tau_\obs-r(z)$, with $r(z)=\int_0^z dz'/(H(z') a_\obs)$, the unperturbed affine parameter $\lambda=r$
and S denotes the source. 
$\lapl_\Om$ is the angular part of the Laplacian in spherical coordinates. Detailed derivations of (\ref{Dez})
can be for instance found in \cite{YFZ09,BD11,CL11}.

The Newtonian approximation of gravity is void of the light cone concept. Moreover, the gravitational potential
couples only to massive bodies, in particular, there is no coupling to light rays. In other words,
light rays cannot probe $\phi_\Newt$. The assumption that there is a light cone embedded in the background cosmology,
albeit an artificial point of view, allows to obtain the first and second term of \eqref{Dez}
in the Newtonian approximation, $\De_\Newt(\bn,z)= \delta_\Newt-\HH^{-1}\dd_r^2 v_\Newt $.
The second term is known as {\it Kaiser-effect} \cite{K87} and is the dominant 
redshift space distortion for small redshifts. Moreover, let us assume that light couples to $\phi_\Newt$
such that its bending around the Sun conforms to actual observations. Including the lensing contribution, 
\be
\De_\Newt(\bn,z)= \delta_\Newt-\HH^{-2}\dd_r^2 v_\Newt -\frac{1}{r_S}\int_0^{r_S}\hspace{-3mm}d\lambda \frac{r_S-r}{r}\lapl_\Om \phi_\Newt  \; .
\label{DezNewt}
\ee
Depending on scale, redshift and redshift binning, the lensing contribution can be the leading redshift space distortion,
which can even dominate over $\delta_\Newt$ \cite{CL11, BD11} for sufficiently distant sources. 

For a more transparent treatment, let us define a Newtonian observable $\De_\Newt(\bn,z)$
through the following replacements in the relativistic quantity \eqref{Dez}: 
\be
\De_\Newt(\bn,z)\equiv \De(\bn,z)\Big|_{\text{D}\rightarrow \de_\Newt, \text{V}\rightarrow v_\Newt, \upPhi=\upPsi \rightarrow \phi_\Newt} \;.\label{DezNewt2}
\ee
Using the results from the last section it follows that $\De_\Newt(\bn,z)=\De(\bn,z)$. 
As a consequence, {\it N-body simulations can be used directly to extract relativistic observables} 
when scalar dust fluctuations on a $\Lambdaup$CDM background are considered at the linear level.

Let us comment on why fluctuations $\delta^H_\NM$ in the Hubble parameter $\bar K=-3H$,
\bea
\delta^H &\equiv& -\frac{\de K/3}{\HH/a} \nonumber \\
&=&\frac{-1}{\HH}\left(\psi ' +\HH\phi + 1/3\, \lapl (w-h')\right) \; ,\label{deH}
\eea
are, in fact, strongly contributing only to the lensing term and, therefore, were identified correctly in \cite{FS12}
as a major source of redshift space distortions. 
Clearly, large fluctuations in the Hubble parameter do not imply that Newtonian N-body simulations 
cannot be trusted. Instead it implies that either \eqref{DezNewt} or \eqref{DezNewt2} (or, better,
an expression including nonlinear effects) should be used to compare numerical experiments   
based on the Newtonian approximation to observations, which was well known
\cite{G67,KS87,YFZ09,Y10,M00,BD11,CL11} before the work \cite{FS12}.

We could have argued based on gauge invariance alone \cite{KS87,BD11} that $\De(\bn,z)$ is constructed
from extrinsic curvature $\delta^H$, intrinsic curvature $R^{(3)} = 4\lapl \psi/ a^2$,
anisotropic extrinsic curvature 
$A_{ij}=a(\partial_i \partial_j/\lapl- \de_{ij}/3) \lapl (w-h')$, and the divergence of the observer's
coordinate acceleration $\vec{a}=\vec{\nabla}\ln[a(1+\phi)]$ in such a way that arguing about
the size of $\delta^H$ adapted to various gauges is meaningless. In certain gauges
$\delta^H$ might qualify as large (NM and S), while in others it qualifies as small (L),
but this does not matter at all. 

Any change in $\delta^H$ induced by a transition between coordinate systems
is compensated for by corresponding changes in $\de^R\equiv \lapl \psi$, $\de^\vec{a} \equiv \lapl \phi$ and $\de^A\equiv \lapl (w-h')$.
This can be checked explicitly for the gauge invariant lensing term 
$\lapl (\upPhi + \upPsi)$, which can be deconstructed into the various gauge dependent 
curvatures as follows:
\be 
\lapl (\upPhi + \upPsi) = \de^R + \delta^\vec{a} + {\de^A}' \;.
\ee 
Note that $\delta^H$ contains the anisotropic extrinsic curvature $\de^A$. 
In the Newtonian matter gauge, $\delta^H_\NM$ becomes large just because $\de^A_\NM$ is large, while in the longitudinal gauge $\de^A_\Long=0$ and $\phi_\Long$, $\psi_\Long$ are quasi-static, meaning they basically remain at their initially
small values (except close to neutron stars and black holes, see \cite{GW11,HW98} for more details).
As a consequence, $\delta^H_\Long$ is negligible. Since $\phi_\Long=\upPhi$ and $\psi_\Long=\upPsi$, all other terms
in \eqref{Dez} involving $\upPsi$ and $\upPhi$, as well as their conformal time derivatives remain much smaller
than the first three terms in \eqref{Dez}. It can be shown that whenever $\de^H_\NM$ contributes to these less relevant terms,
its $\de^A_\NM$ component is either compensated for by another $\de^A_\NM$ contribution or rendered 
harmless by $\lapl^{-1}$. 

The discussion outlined for $\De(\bn,z)$ applies quite generally to any observable.
A constructive algorithm for an arbitrary observable is the following: 
({\it i}) construct the general relativistic gauge invariant observable and express it in terms of (D,V) and $\upPhi$.  
({\it ii}) Use the quasi-static evolution of $\upPhi$ to determine which contributions qualify as large.
If one can identify large contributions that are not reflected in the corresponding Newtonian observable,
then these contributions will have a genuine relativistic origin.
({\it iii})  Employ the dictionary 
$(\text{D},\text{V},\upPhi)\rightarrow (\de_\Newt,v_\Newt,\phi_\Newt)$
to extract relativistic observables using Newtonian N-body simulations. This has been worked out in much greater in detail in \cite{CZ11, GW12}.

\section{conclusion}
In summary, we have shown that fluctuations in the Hubble parameter do not give rise to new, 
so far overlooked redshift space distortions. They contribute to redshift space distortions,
however only to the well-known lensing magnification. 
Therefore contrary to the claims of \cite{FS12}, Newtonian N-body simulations {\it are} the appropriate numerical experiments to extract information
on relativistic observables at least in $1$st order perturbation theory.

\acknowledgements 
It is a great pleasure to thank Dominik Schwarz and Florian Niedermann for most valuable discussions. The work
of TH, SH \& MK was supported by the DFG cluster of excellence 'Origin and 
Structure of the Universe'. The work of SH was supported by TR33 'The Dark Universe'.
\bibliography{DJbib}
\end{document}